\documentclass[conference]{IEEEtran}
\usepackage{cite}

\usepackage{graphicx}
\usepackage[cmex10]{amsmath}
\usepackage[tight,footnotesize]{subfigure}

\usepackage{enumerate}
\begin{document}
%
\title{Power-efficient Hierarchical Data Aggregation using Compressive Sensing in WSN}

\author{\IEEEauthorblockN{Xi Xu}
\IEEEauthorblockA{Department of Electrical and\\Computer Engineering\\
University of Illinois at Chicago\\
Chicago,Illinois,60607\\
Email:xxu25@uic.edu}
\and
\IEEEauthorblockN{Rashid Ansari}
\IEEEauthorblockA{Department of Electrical and\\Computer Engineering\\
University of Illinois at Chicago\\
Chicago,Illinois,60607\\
Email: ransari@uic.edu}
\and
\IEEEauthorblockN{Ashfaq Khokhar}
\IEEEauthorblockA{Department of Electrical and\\Computer Engineering\\
University of Illinois at Chicago\\
Chicago,Illinois,60607\\
Email:ashfaq@uic.edu}}


\maketitle

\begin{abstract}


Compressive Sensing (CS) method is a burgeoning technique being applied 
to diverse areas including wireless sensor networks (WSNs). 
In WSNs, it has been studied in the context of data gathering and aggregation, particularly aimed at
reducing data transmission cost and improving power efficiency. 
Existing CS based data gathering work \cite{BalCS} \cite{HybridCS1} \cite{ HybridCS2} 
in WSNs assume fixed and uniform compression threshold 
across the network, regardless of the data field characteristics.
 In this paper, we present a novel data aggregation architecture model 
 that combines a multi-resolution structure with compressed sensing. 
 The compression thresholds vary over the aggregation hierarchy,
 reflecting the underlying data field. Compared with previous relevant work, 
 the proposed model shows its significant energy saving from theoretical analysis. 
 We have also implemented the proposed CS-based data aggregation framework 
 on a SIDnet SWANS platform, discrete event simulator commonly used for WSN simulations. 
 Our experiments show substantial energy savings, ranging from 37\% to 77\% 
for different nodes in the networking depending on the position of hierarchy.
\end{abstract}
\IEEEpeerreviewmaketitle

\section{Introduction}
Data aggregation \cite{DataAgg} is a key task performed within sensor networks
to fuse information from multiple sensors and deliver
it to a sink node in a manner that eliminates redundancy and enables energy saving.
The redundancy is a consequence of the correlation inherent in smooth data fields, 
such as temperature, pressure, and sound measurements, in practical applications that include surveillance
and habitat monitoring.
Suppose a sensor transfers its single measurement to the sink over $N-1$ intermediate sensors
along a routing path. Each intermediate sensor combines the data it receives with its own data
and forwards it along the route.
This data aggregation process usually involves $O(N^2)$ data transmission.
However when there is redundancy in data then it lends
itself to sparse representations and in-network
compression thereby yielding energy savings in the information
transfer.

Recently the use of compressive data gathering has been examined~\cite{BalCS},
and shown to reduce transmission requirements to $O(N*M)$, where M represents the number of
random measurements, and $M<<N$.
If we ignore temporal change and only consider data in a certain time snapshot,
then each sensor only has one measurement. According to compressive sensing (CS) theory \cite{CS},
the Compressive Data Gathering (CDG) method  \cite{BalCS} requires all the sensors to collectively provide at the sink
at least $M=K \log{N}$ measurements to fully recover the signal, where $K$ is the sparsity of signal.
We note that when the CDG method is applied in a large scale network, $M$ may still be a large number.
Moreover in the initial data aggregation phase in \cite{BalCS}, leaf nodes
unnecessarily transmit $M$ measurements, which is in excess of their sensed data and therefore introduces redundancy in data aggregation. 
Recognizing this, the hybrid CS aggregation\cite{HybridCS1}\cite{HybridCS2} method proposed an amalgam of
the merits of non-CS aggregation and plain CS aggregation.
It optimized the data aggregation cost by setting a threshold $M$
and applying CS aggregation when data gathered in a sensor equals or exceeds $M$.
The data transmission cost and hence the energy consumption is reduced. 
However, we observe that only a small fraction of sensors utilize CS aggregation method
and the transmission measurement number $M = K \log{N} $ for those nodes using CS aggregation method is still large.

Our work here shows significant improvement is possible and it stems from
a hierarchical clustering architecture that we propose. The central
idea is to configure sensor nodes such that instead of one sink
node being targeted by all sensors, several nodes are designated for
intermediate data collection and concatenated to yield a hierarchy
of clusters at different levels. The use of the hierarchical
architecture reduces the measurement number
in the algorithm~\cite{HybridCS1}\cite{HybridCS2} for CS aggregation
since in the new architecture it
is based on the cluster size rather than the global sensor network size $N$.
In this paper, we propose a novel CS-based data aggregation hierarchical architecture over the sensor network
and investigate its performance in terms of data rate and energy savings.
To the best of our knowledge, we are the first to investigate compressive sensing method for hierarchical data aggregation in sensor network.
We refer to our method as Hierarchical Data Aggregation using Compressive Sensing (HDACS).

The proposed data aggregation architecture distributes the workload of one sink to all the sensors, 
which is crucial for balancing energy consumption over the whole network. In this paper we also perform a theoretical analysis of the
data transmission requirements and energy consumption in HDACS.
We implement our proposed architecture on a SIDnet-SWANS simulation platform 
and test different sizes of two-dimensional randomly deployed sensor network. 
The results validate our theoretical analysis. 
Substantial energy savings are reported for a large portion of sensors
on the different hierarchical positions, 
ranging from 50\% to 77\% when compared with~\cite{BalCS}, and from 37\% to 70\%
when compared with \cite{HybridCS1} \cite{HybridCS2}.    

\section{Proposed Data Aggregation Architecture}
\subsection{Model and analysis}
The main idea behind this new architecture is that all sensors will no longer aim at flowing
 their data into one sink. Instead, plenty of collecting clusters have been concatenated 
 forming different types of clusters in different levels. 
 The data flows from the source node through the architecture to the sink. 

Suppose N sensors have been uniformly and randomly deployed in a 2D square space with area
S. Let $s$ be the unit area in the lowest level and the clusters have been defined in a multi-resolution
way with highest level T. In each level $i$, we define:
\begin{itemize}
\item $s_i^{(l)}$: the area of $l^{th}$ cluster
\item $c_i^{(l)}$: the $l^{th}$ cluster head in the network
\item $N_i^{(l)}$: the number of sensors in one cluster 
\item $M_i^{(l)}$: the transmission number of measurements for $l^{th}$ cluster
\item $d_i^{(l)}$:  the sum of distances between cluster head $c_{i}^{(l)}$ 
and its children nodes.
\item  $\gamma_i^{(l)}$: the ratio of transmitting data size and receiving data size.
\item $E_i^{(l)}$: the transmission energy cost in the $l^{th}$ cluster 
\item $C_i$: the collection of cluster heads
\item $ |C_{i}|$ :the number of cluster heads
\item $M_i$: the sum of measurements for transmission within network
\item  $E_i$: the transmission energy cost for all the clusters
\end{itemize}

Here, $C_i$ is defined as the collection of all the clusters, which implies
$C_i=\{c_i^{(1)},c_i^{(2)},\cdots,c_i^{(|C_{i}|)}\}$.  
Total transmission measurements $M_i$ is $M_i=\sum_{l=1}^{|C_i|} M_i^{(l)}$ 
and total energy cost $E_i$ is $E_i=\sum_{l=1}^{|C_i|} E_i^{(l)}$.
\begin{figure}[!t]
\centering
\includegraphics[width=2.5in]{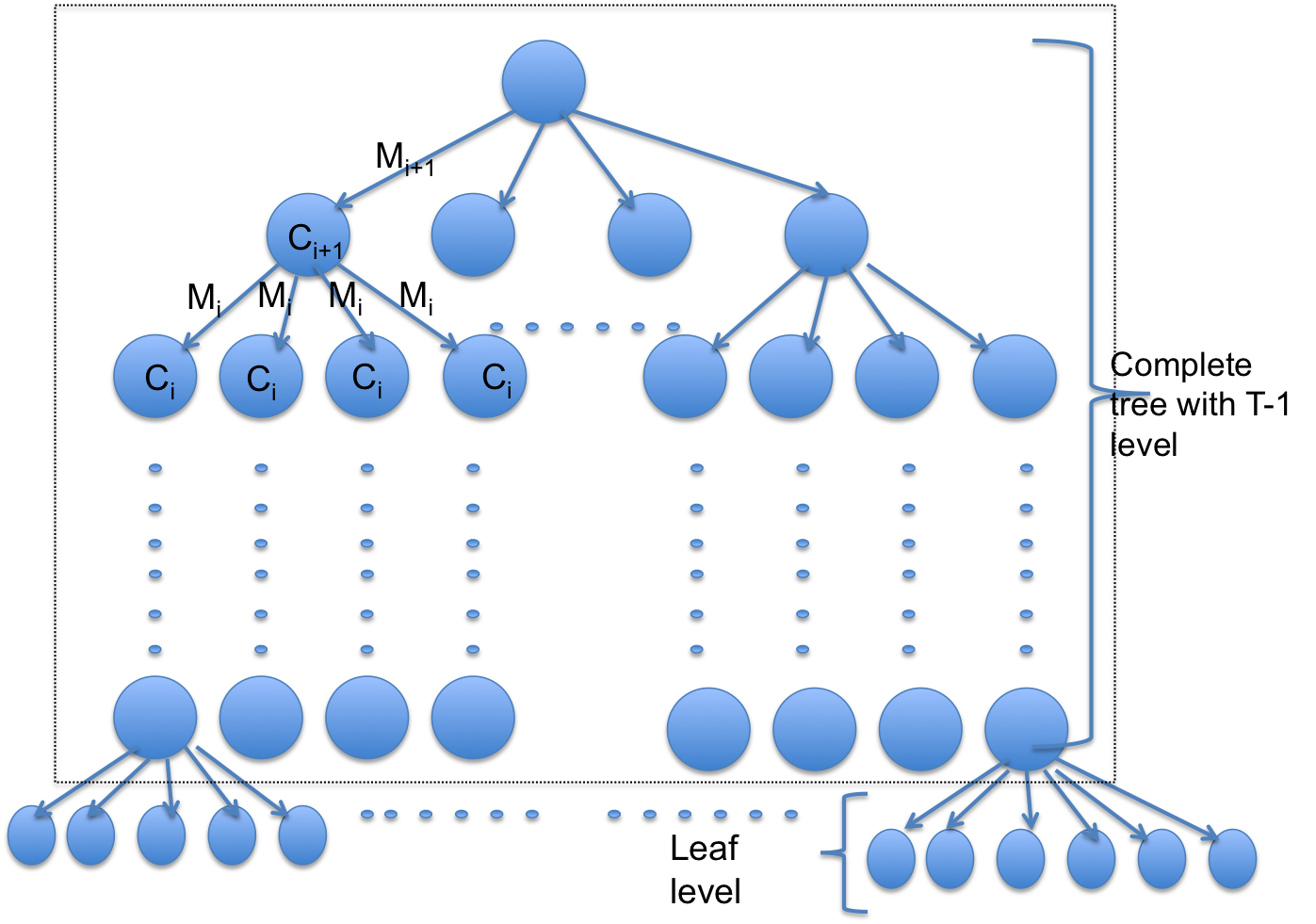}
\caption{CS data aggregation Architecture}
\label{dataAggArc}
\end{figure}

\subsubsection{2D uniformly and randomly network deployment and analysis }
In order to simplify the model analysis and get quantitative comparisons with previous work,
 we put some constraints on our CS data aggregation architecture. Figure \ref{dataAggArc}
 shows the logical tree of clustering configuration, which consists of identical $n$ nodes in level i for $i \geq 2$
and random leaf nodes $N_1^{(l)} \geq n$.
Consider a N nodes network,
where  $N = N' + n^ T$. Therefore, we have the following formula:
\begin{eqnarray*}
  N_1^{(1)}+N_1^{(2)}+N_1^{(3)}+\cdots+N_1^{(|c_1|)}=N\\
  N_1^{'(1)}+N_1^{'(2)}+N_1^{'(3)}+\cdots+N_1^{'(|c_1|)}=N' 
\end{eqnarray*}
Besides, in level $i$, $d_i^{(l)}$ is the sum of distances between cluster head $c_{i}^{(l)}$ 
and its children cluster heads $c_{i-1}^{((l-1)*n+1)}, c_{i-1}^{((l-1)*n+2)}, \cdots, c_{i-1}^{((l-1)*n+n)}$ in the level
i-1. The number of cluster heads is $|C_i|= n^{T-i}$. The area of $l^{th}$ cluster
$s_i^{(l)}$ will be the same as that of all the other clusters at that level. We denote them as $s_i$
which combines $n$ subregions from i-1 level, and it satisfies the relation of
$s_i=n*s_{i-1}$.

We distribute the sensors in a 2D randomly deployed network with some constraints.
\begin{itemize}
\item There will be at least n nodes in each cluster in level 1. This property
requires we have to maximize the probability of n nodes in one cluster:
\begin{equation}
\text{ max } P_1= \frac{n|C_1|}{N}=\frac{n^T}{N}=\frac{1}{1+N'/n^T}
\label{p1}
\end{equation}
It requires to minimize $ N'/n^T$.
From historical experiences, we are prone to set up $n=4^\alpha \text{ for } \alpha\in N^+$
to guarantee the full coverage of the whole region for each clusters with square area without producing intersections
between two neighboring clusters. Therefore, $T=\lfloor log_n^N \rfloor$. And we get the minimum
of $ N'/n^T$.

\item  The remaining sensors are uniformly and randomly distributed
in $|C_1|$ clusters. So we set to maximize the probability of of $N'$ nodes in
$|C_1|$ clusters.
\begin{equation}
\text { max } P_2= \frac{|C_1|}{N'}= \frac{n^{T-1}}{N'} 
\label{p2}
\end{equation}
This has already been achieved in the constrains (\ref{p1}).
\end{itemize}

The main advantage of this network deployment is that it is based on 2D randomly 
deployed network topology, which corresponds to practical sensors
distribution. It also addresses issues when the 
the condition that $N=n^T$ is not met. Besides, the number of cluster heads will be at most 
$n^{T-1}$, and the leaf nodes will be $N-n^{T-1} \geq n^T-n^{T-1} =(n-1)
n^{T-1}$. If $n>2$, $N-n^{T-1}>n^{T-1}$. This result implies that only a small
number of nodes will be involved with multiple level data processing and
aggregation. The only job of other sensors is just sending their data
directly to the cluster head. The balance in load distribution is achieved by randomly 
choosing different cluster heads in each duty circle.
 
\subsubsection{The process of CS data aggregation}

In the initial phase, $N_1^{(l)}-1$ sensors in each 
region only send their raw data to their cluster head $c_1^{(l)}$, which adopts the same strategy
as paper \cite{HybridCS1} \cite{HybridCS2} so as to reduce CS data aggregation redundancy. 
$c_1^{(l)}$ compressed them into $M_1=K*\log{N_1^{(l)}}$ random
measurements. In level i ( $ i \geq 2 $),  the $l^{th}$
cluster head $c_i^{(l)}$ receives $M_{i-1}^{(j)}$ random measurements, where $j \in[(l-1)*n+1, (l-1)*n+n]$
 from its children cluster head $c_{i-1}^{(j)}$. It performs CS recovery algorithm to
reconstruct the redundant data. 
After accumulating all the data from their children nodes,
the cluster head takes $M_i^{(l)}$ random measurements of the signal and send
them to its parent cluster head in the level i+1.
According to compressive sensing theory, for signal with sparsity K, the random measurements $O(K\log{N})$ will
be enough to fully represent original signal with cardinality N. 
We adopt the method in \cite{HybridCS1} \cite{HybridCS2} to set up threshold
$M= K*\log N$ to optimize the data transmission size. 
We propose to set up multiple thresholds $M_i^{(l)}=K*\log {N_i^{(l)}}$ with upper bound
 $O(K\log{N})$ in the top level T.  $M_i$ is a small number when i is small.
 This property helps to reduce the data transmission number and hence
 significantly save energy. 
\subsubsection{Parameters Analysis}
\begin{figure}[!t]
\centering
\includegraphics[width=2in]{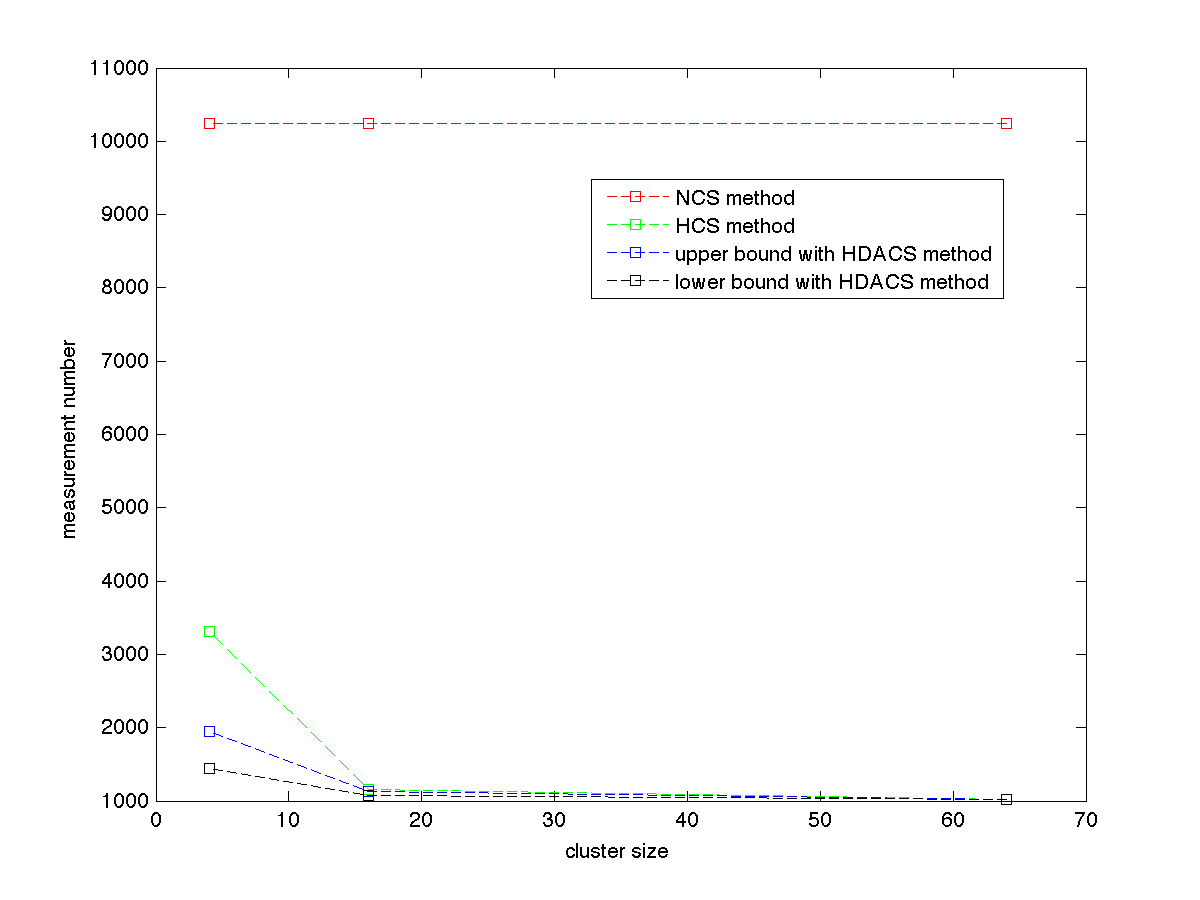}
\caption{Measurements Comparison for 1024 Sensor Networks }
\label{MeaCom}
\end{figure}

\begin{figure}
\centering
\subfigure[Cluster Size 4  ]{\includegraphics[width=2in]{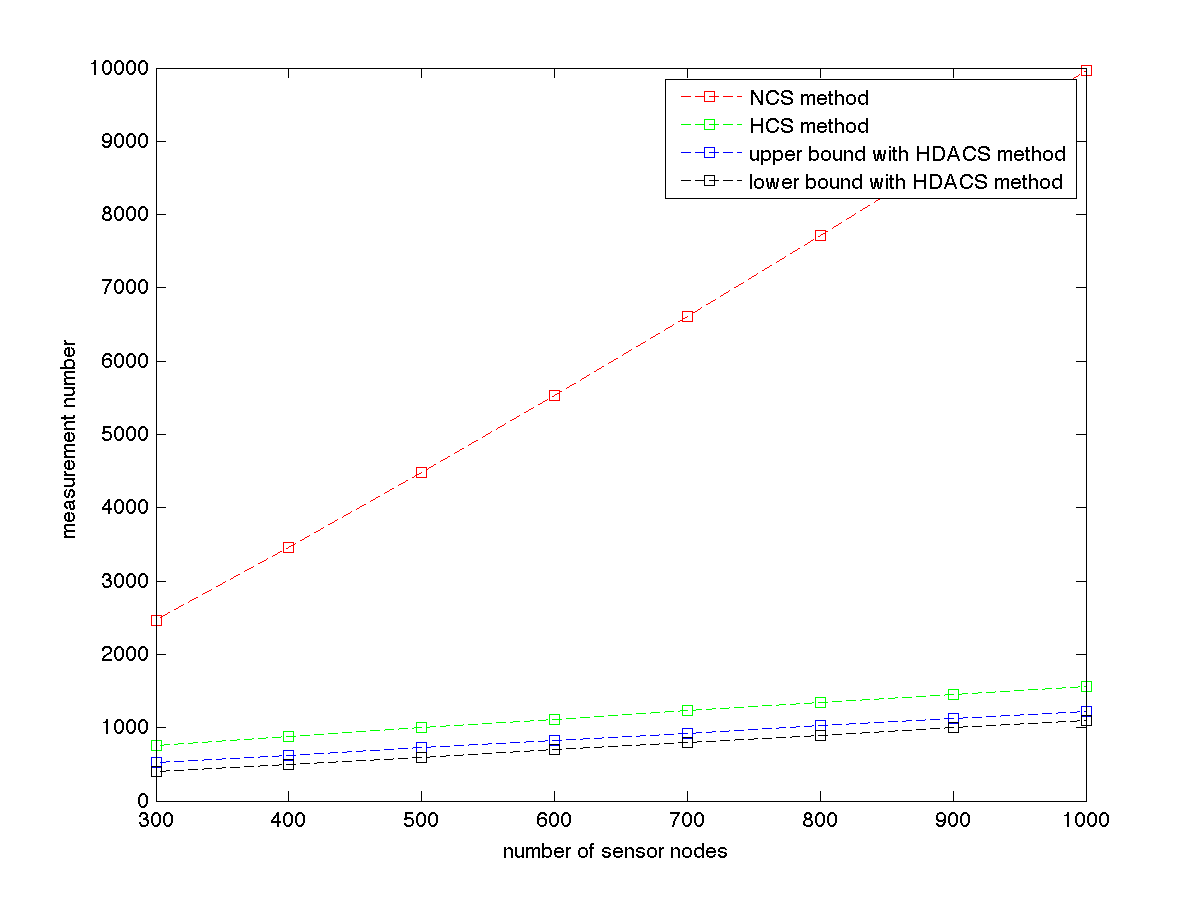}}
\hfil
\subfigure[Cluster Size 16]{\includegraphics[width=2in]{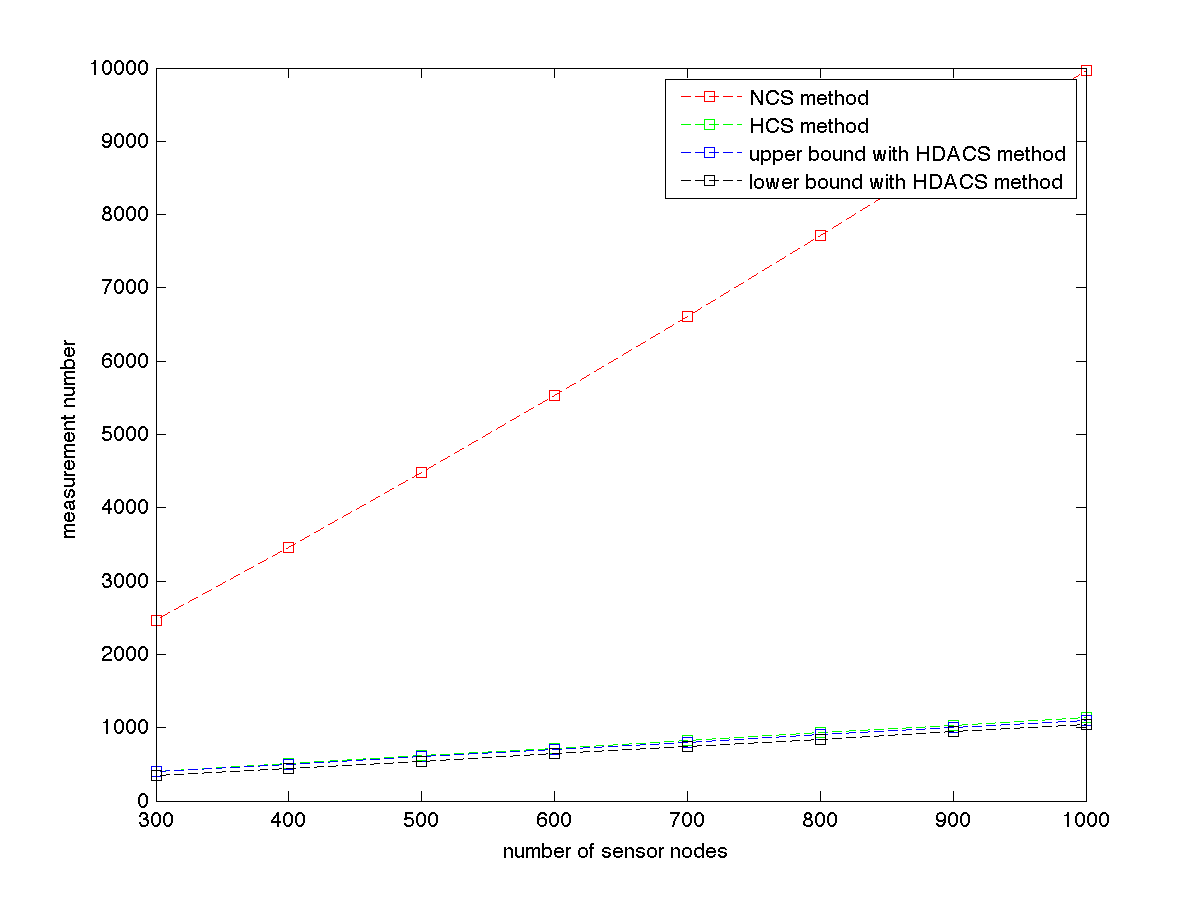}}
\caption{Measurement Comparison for Different Sizes of Sensor Networks}
\label{diffSize}
\end{figure}

For $ n^{T+1}\geq N=n^T+N'\geq n^{T} $,
we get $n^{i+1}\geq N_i^{(l)} \geq n^i $ and measurements $M_i^{(l)}$ in the range of $[iK\log{n},
(i+1)K\log{n}]$.
The total transmission measurements $M$ for the whole data aggregation task is:
%
\begingroup
\fontsize{8pt}{12pt}\selectfont
\begin{eqnarray*}
  M &=& \sum_{i=1}^{T} M_i=\sum_{i=1}^{T}\sum_{l=1}^{|C_i|} M_i^{(l)}\\
      &=& \sum_{l=1}^{|C_1|} (N_1^{(l)}-1) + \sum_{i=2}^{T} \sum_{l=1}^{|C_i|} (n-1) M_{i-1}^{(l)} \\
\end{eqnarray*}
\endgroup
Let $S_1=\sum_{i=1}^{T-1} \frac {i}{n^i} $ and and $S_2= \sum_{i=1}^{T-1} \frac {1}{n^i}$we get the closed form of $S_1$:
\begingroup
\fontsize{8pt}{12pt}\selectfont
\[
 S_1 =\frac{1/n(1-1/n^{T-1})}{(1-1/n)^2}-\frac{T-1}{n^T(1-1/n)}\\
\]
\endgroup
and 
\begingroup
\fontsize{8pt}{12pt}\selectfont
\[
 S_2 =\frac{1/n*(1-1/n^{T-1})}{1-1/n}
\]
\endgroup
Therefore, the lower bound of data transmission number M is:
\begingroup
\fontsize{8pt}{12pt}\selectfont
\[
  \Omega(M) = N-n^{T-1}+ K(n-1) n^{T-1}\log {n} S_1
\]
\endgroup
and upper bound is 
\begingroup
\fontsize{8pt}{12pt}\selectfont
\[
  O(M) = N-n^{T-1}+ K(n-1) n^{T-1}\log {n} (S_1+S_2)
\]
\endgroup
On the other hand if data is sent using the same data architecture, 
the total measurements with the plain or non-hybrid CS (NCS) algorithm in paper \cite{BalCS} is:
$M_{NCS}=N*K*\log{N}$. In paper \cite{HybridCS1} \cite{HybridCS2}, 
the total measurements for hybrid CS (HCS) algorithm is:
$M_{HCS}=\sum_{l=1}^{|C_1|} (N_1^{(l)}-1) + \sum_{i=2}^{T} \sum_{l=1}^{|C_i|} (n-1) K\log N
= N-n^{T-1}+K(n-1)n^{T-1} \log{N} S_2$.
In the following analysis, we assume the sparsity K as unity to
rule out the effects from data field for data aggregation comparison. 
Figure \ref{MeaCom} shows the quantitative comparison of total data transmission measurements with 
cluster size n = 4, 16, 64 for proposed HDACS method, NCS
data aggregation \cite{BalCS} and HCS method \cite{HybridCS1} \cite{HybridCS2} with 1024 sensor nodes. 
From figure \ref{MeaCom}, we find the bigger the cluster size is, the less
measurements needed for data transmission. However, this theoretical analysis
does not consider the realistic routing protocol underlying the network
architecture in the lower layer.
Simply expanding the cluster size within local cluster and all the nodes forward their sensed data into cluster head directly, which
definitely will lead to severe data flooding and data loss. Therefore, cluster size will be fixed as 4 and 16
in the following analysis. Figure \ref{diffSize} shows total data transmission measurements 
changes with increase of sensor nodes under these two fixed cluster sizes.
From the figure, we observe that the NCS method introduces a large number of data
redundancy. The measurements required by HCS method is a little worse
than proposed method, but we need to point out that this comparison is based on
the premise that the data is propagated on the muli-resolution data
architecture. Since a lot of sensors are leaf nodes and only transmit their raw
data to their cluster heads both in the proposed method in this paper and 
HCS method \cite{HybridCS1}\cite{HybridCS2} in the first level, 
they lead to very similar result in the theoretical analysis. 

The data compression ratio $\gamma_i$ is calculated as follows:
\[
\gamma_i = 
     \begin{cases}
        \frac{M_1^{(l)}}{N_1^{(l)}} =\frac{K\log{N_1^{(l)}}}{N_1^{(l)}} & \text{if }  i=1\\
       \frac{M_i^{(l)}}{\sum_{j=(l-1)n+1}^{(l-1)n+n} M_{i-1}^{j} } & \text{if } i \geq 2 
      \end{cases}
\]

\begin{figure}[!t]
  
\centering

\centering
\subfigure[Cluster Size 4  ]{\includegraphics[width=2in]{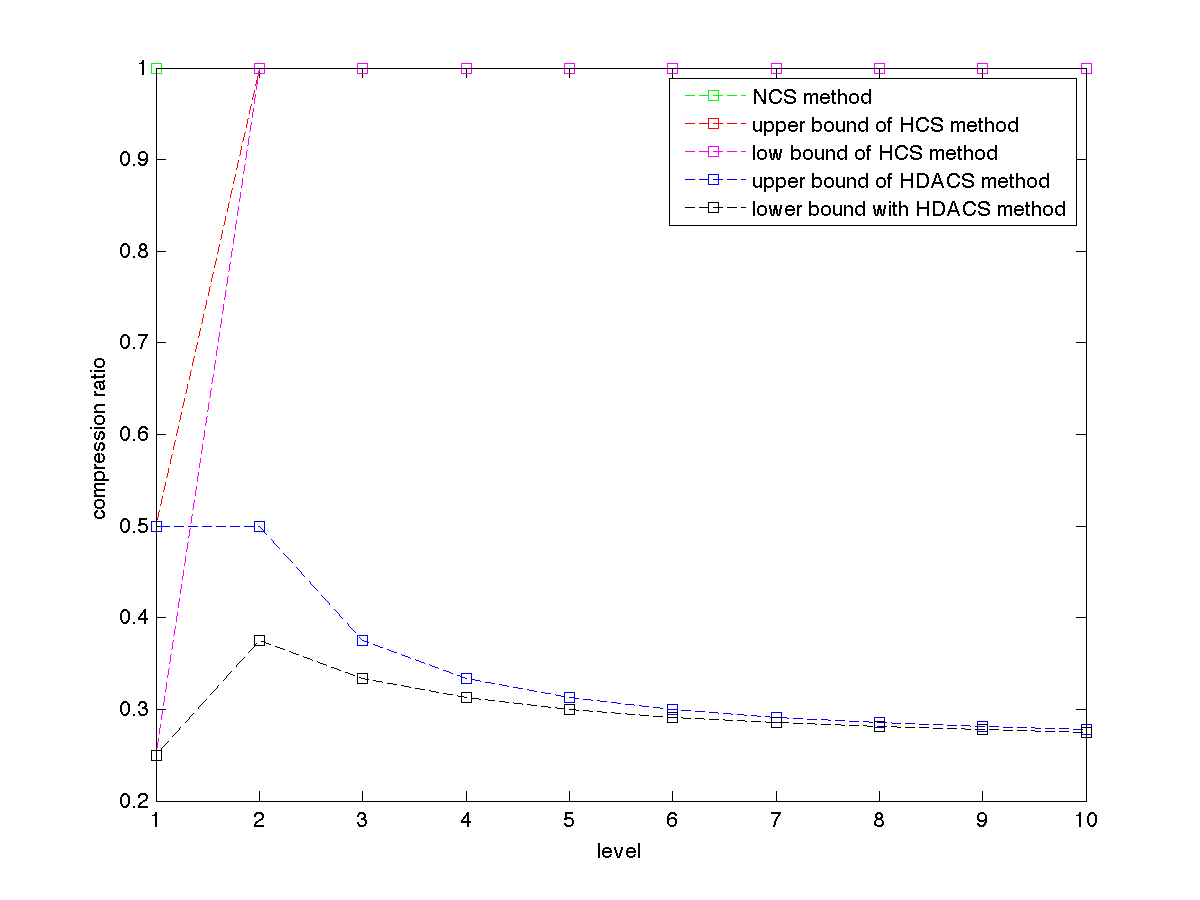}}
\hfil
\subfigure[Cluster Size 16]{\includegraphics[width=2in]{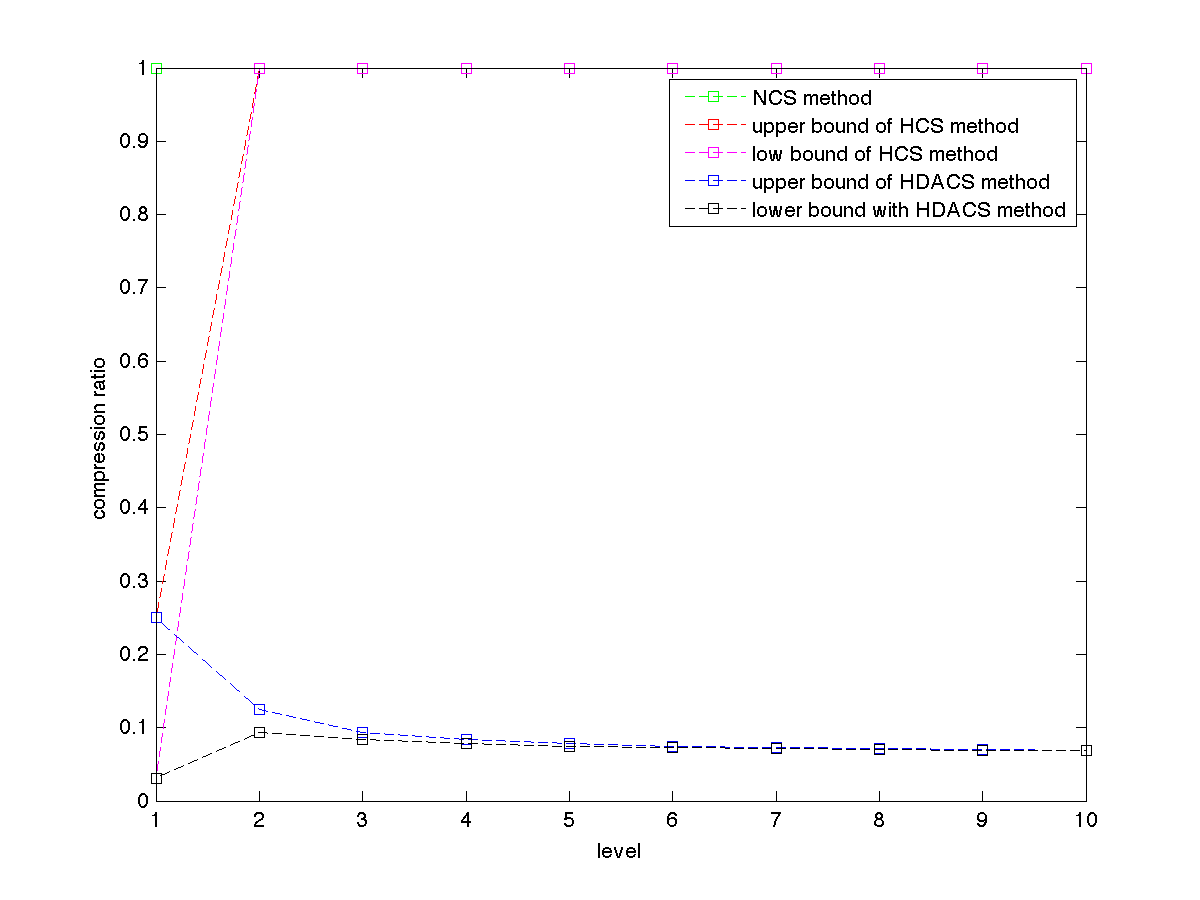}}\\
\caption{Compression ratio change with increase of levels for Different Sizes of Sensor Networks}
\label{CR}
\end{figure}

Compression ratio resides in the range of: 
$\gamma_i \in [\frac{1}{n}(1+\frac{1}{i-1}), \frac{1}{n}(1+\frac{1}{i})], \text{if } i\geq 2$.
Figure \ref{CR} shows the compression ratio changes in each level for NCS
method, HCS method, and proposed HDACS method with cluster size 4 and 16.
The figure shows that the NCS method provides no compression at all. HCS method yields
compression in the first level by using compressive sensing method to
compress data when the number of data reaches the global threshold. The
proposed hierarchy based method achieves a remarkable compression ratio for each
level which is well below 0.5.
This is very appealing as nodes that are spatially close to 
the central sink working as intermediate nodes usually consume more energy than
other nodes.
The energy savings in the task in the application layer for those nodes will balance the 
energy consumed in the routing layer, which therefore prolongs the lifetime of the network. 
 
For the energy analysis, we pay more attention to the data transmission cost
and the cost of receiving data is taken as a constant. 
Transmission energy cost $E_i^{(l)}$ is usually a function of transmission distance $d_i^{(l)}$ and data size $M_i^{(l)}$.
Therefore, $E_i^{(l)}$ has been modeled as $E_i^{(l)} = c_{s}+ cd_i^{(l)} M_i^{(l)}$, 
which leads to the total transmission energy cost in each level $E_i=\sum_{l=1}^{|C_i|} E_i^{(l)}$.
Here, $c_{s}$ is a constant startup energy consumption for each data transmission task, and 
$c$ is a constant transmission cost for unit data size per unit distance. 

Assume $d_i^{(l)} =\sum_{t=1, (x_t,y_t)\in c_i}^{n} [(x_t-x_{c_i})^2+(y_t-y_{c_i})^2]^{1/2} $, where
$ (x_{c_i}, y_{c_i})$ and $(x_t,y_t) $ are the location coordinates of $C_i^{(l)}$ and its children nodes respectively. 
In a large dense uniformly and randomly distributed sensor network, if $i\geq 1$
$d_i^{(l)} = 4 \frac{(n-1)} {s_i} \int_0^{b_i} \int_0^{b_i} \! (x^2+y^2)^{1/2} \, \mathrm{d} x\mathrm{d} y
= \frac{1}{12} \pi n^{\frac{i-1}{2}} s^{1/2} (n-1)$, where $b_i= \frac{1}{2} s_i ^{1/2}$.
And for $i=1$, $d_1^{(l)} = \frac{1}{12} \pi s^{1/2} (N_1^{(l)}-1) $.

The final total energy consumption will be:
\begingroup
\fontsize{8pt}{12pt}\selectfont
\begin{eqnarray*}
E &=&\sum_{i=1}^{T}\sum_{l=1}^{|C_i|} E_i^{(l)} \\
   &=&\sum_{l=1}^{|C_1|} c_s + c d_1^{(l)} + \sum_{i=2}^{T}\sum_{l=1}^{|C_i|} c_{s}+ cd_i^{(l)} M_i^{(l)} \\
   &=& n^{T-1} c_s +c \frac{1}{12} \pi s^{1/2} \sum_{l=1}^{|C_1|} (N_1^{(l)}-1)\\ 
   &+& \sum_{i=2}^{T}n^{T-i} c_s+ \sum_{i=2}^{T} c\frac{1}{12} \pi n^{\frac{i-1}{2}} s^{1/2} (n-1) K\log{N_{i-1}^{(l)}}   \\
   &=& n^{T-1} c_s +c \frac{1}{12} \pi s^{1/2} \sum_{l=1}^{|C_1|} (N_1^{(l)}-1)\\ 
   &+& n^{T-1} c_s S_2 +c\frac{1}{12} \pi s^{1/2} (n-1) K \sum_{i=2}^{T} \sum_{l=1}^{|C_i|} n^{\frac{i-1}{2}}\log{N_{i-1}^{(l)}}     \\
\end{eqnarray*}
\endgroup

Let $S_1'=\sum_{i=1}^{T-1} i n^{-i/2} $ and we get the closed form of $S_1'$:
\begingroup
\fontsize{7pt}{12pt}\selectfont
\[
 S_1' =\frac{n^{-1/2}(1-n^{-(T-1)/2})}{(1-n^{-1/2})^2}-\frac{(T-1)n^{-T/2}}{1-n^{-1/2}}
\]
\endgroup
And $S_2'=\sum_{i=1}^{T-1}=n^{-i/2}$ and its closed form is:
\begingroup
\fontsize{7pt}{12pt}\selectfont
\[
S_2'=\frac{n^{-1/2}(1-n^{-(T-1)/2})}{1-n^{-1/2}}
\]
\endgroup
Therefore, the lower bound of total energy consumption E is:
\begingroup
\fontsize{7pt}{12pt}\selectfont
\[
  \Omega(E)= n^{T-1}c_s(1+S_2) + c \frac{1}{12} \pi s^{1/2}  [(N-n^{T-1})+K(n-1)n^{T-1} S_1'\log{n}]
\]
\endgroup
and upper bound of E is:
\begingroup
\fontsize{7pt}{12pt}\selectfont
\[
   O(E)= n^{T-1}c_s(1+S_2) + c \frac{1}{12} \pi s^{1/2} [(N-n^{T-1})+K(n-1)n^{T-1} (S_1'+S_2') \log{n}]
\]
\endgroup

Follow a similar derivation, we get the transmission energy consumption for NCS
method in paper \cite{BalCS} with the same data aggregation architecture 
\begingroup
\fontsize{7pt}{12pt}\selectfont
\[
E_{NCS}=n^{T-1}c_s(1+S_2)+c \frac{1}{12} \pi s^{1/2}  \log{N}[ (N-n^{T-1})+K (n-1)n^{T-1} S_2']
\]
\endgroup
and energy consumption with HCS method in paper \cite{HybridCS1} \cite{HybridCS2} 
\begingroup
\fontsize{7pt}{12pt}\selectfont
\[
E_{HCS}=n^{T-1}c_s(1+S_2)+c \frac{1}{12} \pi s^{1/2}  [(N-n^{T-1})
+ K(n-1)n^{T-1}\log{N} S_2']
\]
\endgroup

%


\begin{figure}[!t]
\centering
\includegraphics[width=2in]{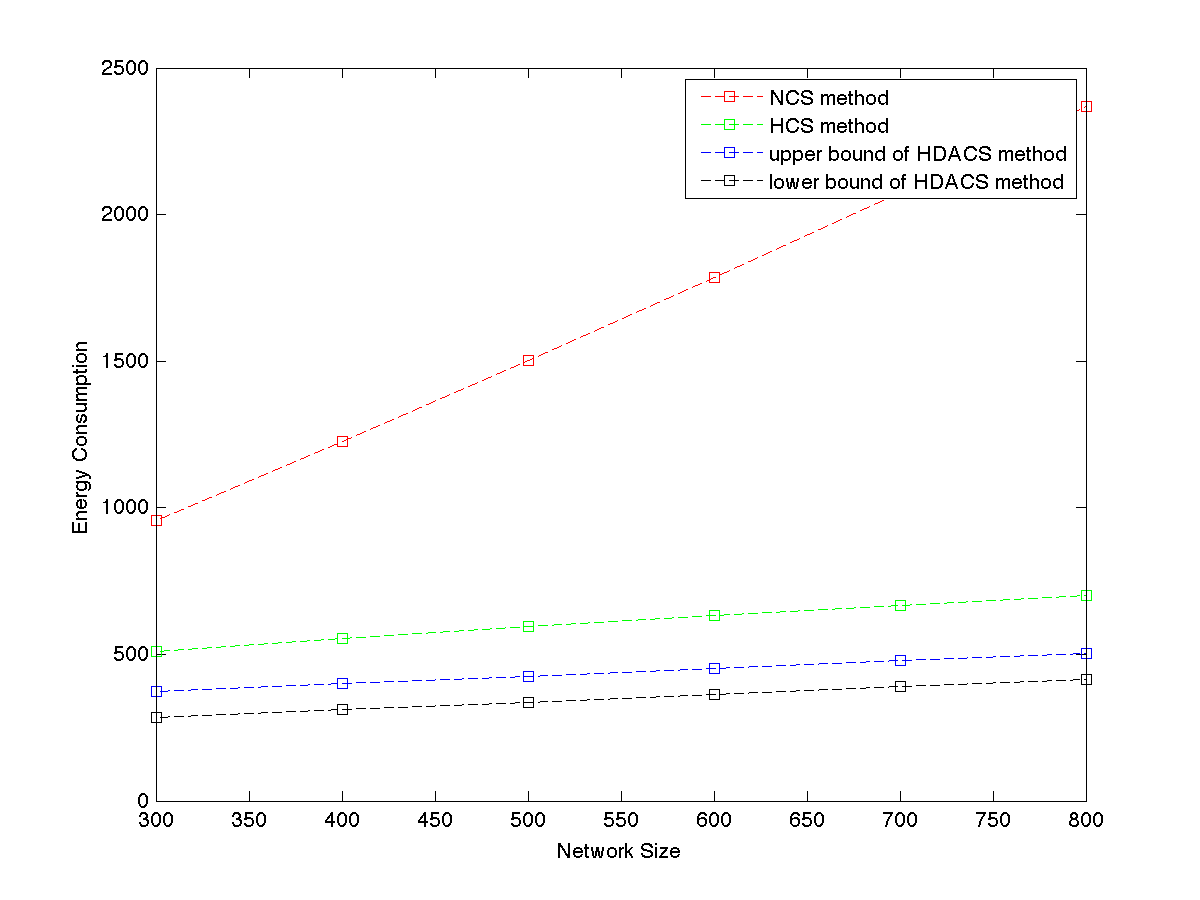}

\caption{Energy consumption change with increase of  network size}
\label{EC_networkSize}
\end{figure}

To ignore the effects of all the constant parameters, we assume $c_s, c, K, s$ as unity.
Figure \ref{EC_networkSize} reflects final transmission energy consumption $E$ trend 
with 300, 400, 500, 600, 700, 800 network scale for cluster size 4.
The proposed HDACS method achieves the highest efficiency in energy consumption 
compared with other methods. 
In the following paper, we set up 2D irregular network deployments on Java-based 
SIDnet-SWANS simulation platform to demonstrate feasibility and robustness of our hierarchy model.

\subsection{Implementation procedure}
\subsubsection{Signal model}

A variety of practical applications in survelillance and habitat monitoring,
the data fields such as temperature, sound, pressure measurements are usually smooth. 
In this paper we ignore the effect of variation of sparsity K in each level.
Therefore, smooth data field with uniform noise is a practical choice to get the
sparse signal representation with identical sparsity K.
We perform Discrete Cosine Transform (DCT) for each of the collecting clusters
before taking random measurements.
The main reasons for choosing DCT are: a). It yields fast vanishing moments of signal representation
and gives real coefficients unlike Discrete Fourier Transform (DFT). b). It also
does not require that cardinality of measurements be a power of 2 as wavelet
transform does.
We perform the truncating process for DCT coefficients by forcing
those magnitudes below a threshold to zero in order to further sparsify the signals. 
The threshold has been set up by
$\alpha$ percentile of the first dominant magnitudes. In actual simulation,
$\alpha$ is chosen as 0.01, 0.005.

\subsubsection{Routing model and recovery algorithm}

The multi-scale routing protocol matches well with hierarchical data aggregation  
mechanism. Since our model mainly focuses on the dense and large-scale network topology, 
it guarantees the existence of shortest path between any two nodes.

CoSaMP algorithm \cite{CoSaMP} has
been adopted as the CS recovery algorithm in our implementation. This algorithm
takes $y=\Phi*\Phi x$ as a proxy to represent signal inspired by the restricted isometry property 
of compressive sensing. Compared with other recovery methods such as various
versions of OMP\cite{OMP1}\cite{OMP2} algorithms, convex programming methods\cite{Convex1}\cite{Convex2}, 
combinatorial algorithms\cite{combinatorial1}\cite{combinatorial2},
CoSaMP algorithm guarantees computation speed and
provides rigorous error bounds.

\subsubsection{Simulation results}
\begin{figure}[!t]
\centering
 \includegraphics[width=2.5in]{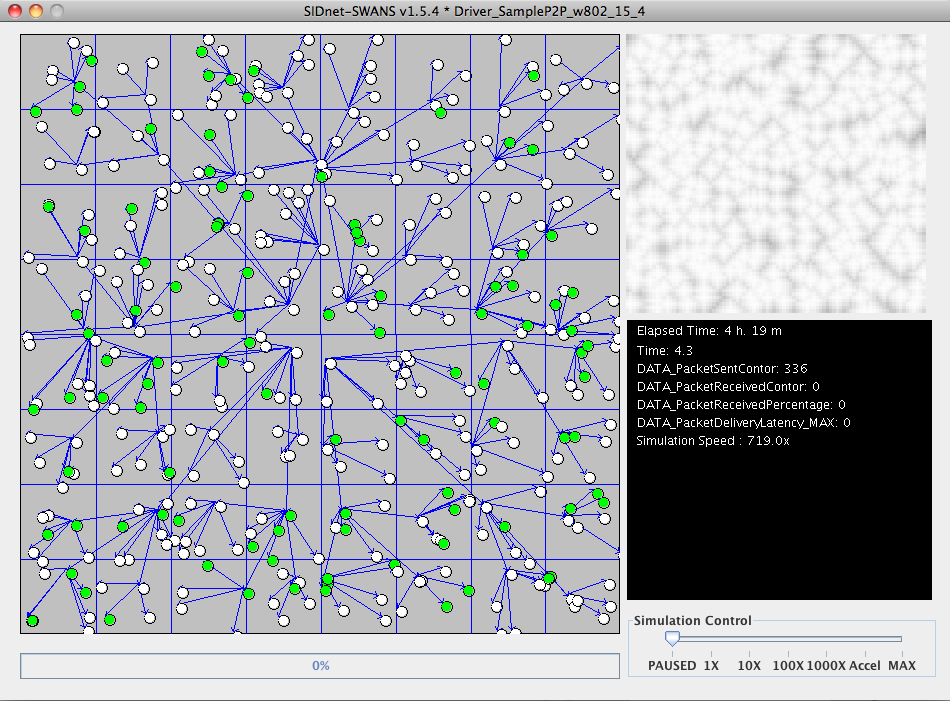}
 \caption{Simulation User Interface on SIDnet-SWANS platform}
\label{simuImage}
\end{figure}

SIDnet-SWANS \cite{SIDnet} is a sensor network simulation environment for various aspects of
applications, which provides with Java based visual tool, has been utilized 
to study the performance of the proposed algorithm. 
The JiST system, which stands for Java in Simulation Time, is a Java-based discrete-event simulation
 engine.  JiST system has been used to obtain the transmission time and energy consumption for each
 sensor. Figure \ref{simuImage} is a snapshot of user interface of newly designed CS data
aggregation architecture on SIDnet-SWANS for 400 sensors network. 
 In this section the performance has been evaluated on SIDnet-SWANS platform with JiST system to 
 demonstrate all the theoretical analysis process .  
 
\begin{figure}[!t]
\centering
\includegraphics[width=2.5in]{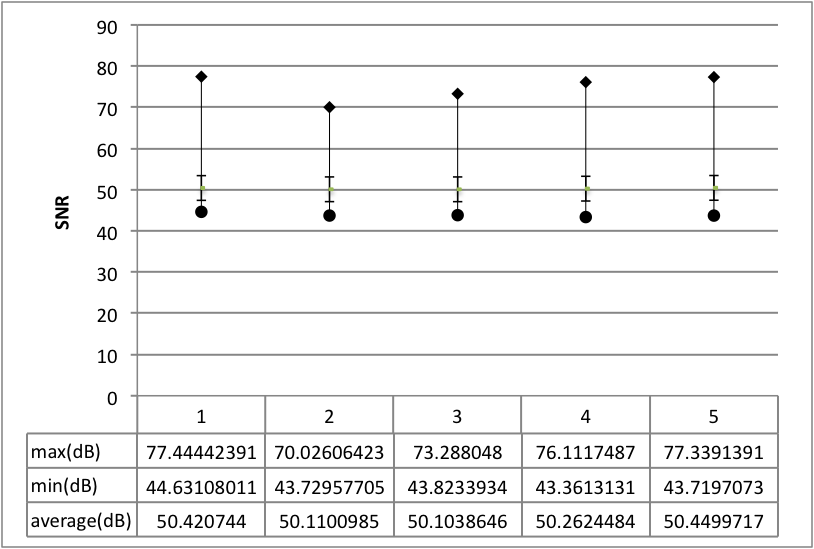}
\caption{Recovery performance for 300,400,500,600,700 sensors}
\label{SNR}
\end{figure}

 The algorithms was tested against five network sizes: 300, 400, 500, 600, and 700 nodes over 
 a flat data field with uniformly distributed additive white noise. In all these network, we choose $n=4$ and $T=3$. 
 The leaf nodes number $N_1$ in the level one is flexible, which fits the characteristics of
 two-dimensional random deployment of sensor networks. Therefore, $N = N_1 + n^
 {T-1}$.
 In the recovery procedure, we adopt the idea of Model-based CoSaMP \cite{ModelbasedCOSAMP} algorithm
as the DCT representation makes the support location of coefficients visible
and design a new CoSaMP algorithm for DCT based signal ensemble, which accurately
recovers the data. We define the signal to noise ratio (SNR) as the logarithm of
the ratio of signal power from each sensors over recovery error in the fusion center. 
 As we see from figure \ref{SNR}, the change of sensor size does not affect SNR performance. 
 
\begin{figure}[!t]
\centering
\includegraphics[width=2.5in]{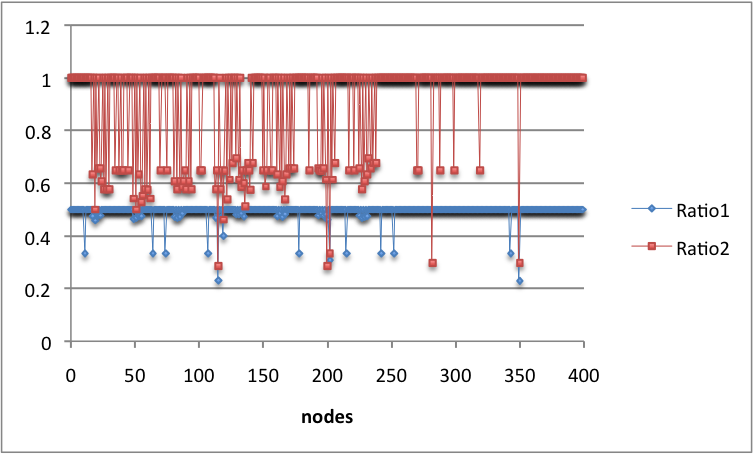}
\caption{Transmission energy consumption ratio for 400 sensors. 
Ratio1 is defined as transmission energy consumption ratio of proposed HDACS and NCS data
aggregation. Ratio2 is defined as transmission energy consumption ratio of proposed HDACS and HCS data aggregation }
\label{ER}
\end{figure}

Figure \ref{ER} shows the comparison of transmission energy consumption distribution for 400 sensor networks.
 Ratio1 is defined as transmission energy consumption ratio
 of proposed HDACS and NCS data aggregation.
Ratio2 is defined as transmission energy consumption ratio of proposed HDACS and HCS data aggregation.
As we see from the figure \ref{ER}, Ratio1 is
less than 0.5, which means 50\% transmission energy will be saved compared with NCS data aggregation.
Ratio2 is almost equal or less than 1, which is owing to the fact that most nodes only transmit data in
the level one and finish their job. Both proposed HDACS and HCS data aggregation
 adopt the same strategy that only raw data is transmitted
for those leaf nodes, which explains why most Ratio2 values of nodes are equal to one. But for those nodes working as
collecting clusters in the levels that are higher than one, Ratio2 values are less or equal to 0.633 as we expect.
The nodes with highest level save almost 70\% power. Moreover, the results we
obtain so far depend on the frame size per transmission in MAC layer to some extent. If the data size becomes larger,
data will be segmented into more frames for transmission. And this will definitely  cost more power.
Since the comparison of proposed HDACS, NCS and HCS algorithms always refers to compare the number of $\log{N_i}$ and $\log{N}$. 
Suppose one frame size is $m$, then the frame number of data size $N_i$ and $N$ are $\left \lceil \frac{\log{N_i}}{m} \right \rceil$ and 
$\left \lceil \frac{\log{N}}{m} \right \rceil$ respectively. If $N_i \approx n^ i$
and two frame number are $\left \lceil \frac{i\log{n}}{m} \right \rceil$ and $\left \lceil \frac{T\log{n}}{m} \right
\rceil$. When $i=2$ , $n=4, m=4$ and $T=4$, frame number are 1 and 2 respectively, which explains how 50\% transmission energy
is saved by using HDACS data aggregation . 

\section{Conclusion and future work}
In this paper, we presented a novel power-efficient hierarchical data aggregation architecture using compressive sensing for 
a large scale dense sensor network. It was aimed at reducing the data aggregation
complexity and therefore enabling energy saving. The proposed architecture is designed by setting up
multiple types of clusters in different levels. The leaf nodes in the lowest level only transmit the
raw data. The collecting clusters in other levels perform DCT to get sparse
signal representation of data from their own and children nodes, take random measurements
and then transmit them to their parent cluster heads. When parent collecting clusters receive random
measurements, they use inverse DCT transformation and DCT model based CoSaMP algorithm to
recover the original data. By repeating these procedures, the cluster heads in the top level 
will collect all the data. We perform theoretical analysis of hierarchical data aggregation
model with respect to total data transmission number, data compression ratio and
transmission energy consumption. We also implement this model on SIDnet-SWANS
simulation platform and test different sizes of two-dimensional randomly
deployed sensor network. The results demonstrate the validation of our model.
It guarantees the accuracy of collecting data from all the sensors. 
The transmission energy is significantly reduced compared with the
previous work. 

In our future work, we will also take into consideration changeable
factors of sparsity $K$. It refers to more complex data fields, and adaptive model
will be set up to handle the dynamic nature of data aggregation fields. 
Besides, other CS recovery algorithms will also be investigated
to reduce recovery complexity and improve signal recovery accuracy. Distributed compressive
sensing\cite{DCS}, that factors in the spatial correlation of
data, turns out to be a very promising recovery algorithm. Moreover, other tasks besides data
aggregation will also be exploited on our proposed hierarchical architecture.

\begin{thebibliography}{20}

\bibitem{DataAgg}
Ramesh Rajagopalan and Pramod K. Varshney, 
\emph{Data aggregation techniques in sensor networks: A survey},
 IEEE Communications Surveys and Tutorials, vol. 8, no. 4, 2006

\bibitem{BalCS}
Chong Luo, Feng Wu, Jun Sun,Chang Wen Chen. \emph{Compressive Data Gathering for Large-Scale 
Wireless Sensor Networks}, MobiCom’09, Beijing, China, September 20–25, 2009. 

\bibitem{CS}
David L. Donoho. \emph{Compressed Sensing}, IEEE TRANSACTIONS ON INFORMATION THEORY, 
VOL. 52, NO. 4, APRIL 2006

\bibitem{HybridCS1}
Jun Luo, Liu Xiang,Catherine Rosenberg,\emph{Does Compressed Sensing Improve the 
Throughput of Wireless Sensor Networks?}, In Proceedings of the IEEE International Conference 
on Communications (ICC’10), pp 1—6, Cape Town, South Africa, May 2010 

\bibitem{HybridCS2}
Liu Xiang, Jun Luo, Athanasios V. Vasilakos, \emph{Compressed data aggregation for
 energy efficient wireless sensor networks}, in Proc. of the 8th IEEE SECON, 2011, pp. 46–54.

 \bibitem{CoSaMP}
D. Needell and J. Tropp, \emph{CoSaMP: Iterative signal recovery from in- complete
 and inaccurate samples}, Appl. Computat. Harmon. Anal., vol. 26, no. 3, pp. 301–321, May 2009. 

\bibitem{OMP1}
J. A. Tropp and A. C. Gilbert. \emph{Signal recovery from random measurements via 
orthogonal matching pursuit}.
IEEE Trans. Info. Theory, 53(12):4655–4666, 2007.

\bibitem{OMP2}
D. L. Donoho, Y. Tsaig, I. Drori, and J.-L. Starck. \emph{Sparse solution of underdetermined linear equations by
stagewise Orthogonal Matching Pursuit (StOMP)}. IEEE Trans. Inf. Theory, 2007.

\bibitem{Convex1}
I. Daubechies, M. Defrise, and C. De Mol.\emph{ An iterative thresholding algorithm for linear inverse problems with
a sparsity constraint}. Comm. Pure Appl. Math., 57:1413–1457, 2004.

\bibitem{Convex2}
M. A. T. Figueiredo, R. D. Nowak, and S. J. Wright. \emph{Gradient projection for sparse reconstruction: 
Application to compressed sensing and other inverse problems}. IEEE J. Selected Topics in Signal Processing: 
Special Issue on Convex Optimization Methods for Signal Processing, 1(4):586–598, 2007.

\bibitem{combinatorial1}
A. Gilbert, M. Strauss, J. Tropp, and R. Vershynin. \emph{Algorithmic linear dimension reduction 
in the l1 norm for sparse vectors}. August 2006.

\bibitem{combinatorial2}
A. Gilbert, M. Strauss, J. Tropp, and R. Vershynin.\emph{ One sketch for all: Fast algorithms
 for compressed sensing}. In Proc. 39th ACM Symp. Theory of Computing, San Diego, June 2007.


\bibitem{ModelbasedCOSAMP}
Richard G. Baraniuk, Volkan Cevher, Marco F. Duarte,and Chinmay Hegde. \emph{Model-Based
 Compressive Sensing}, IEEE TRANSACTIONS ON INFORMATION THEORY, VOL. 56, NO. 4, APRIL 2010

\bibitem{SIDnet} 
Oliviu C. Ghica, \emph{SIDnet-SWANS Manual}, Northwestern University, March 3, 2010

\bibitem{Jist}
RimonBarr, \emph{JiST-JavainSimulationTimeUserGuide}, March19, 2004, barr@cs.cornell.edu

\bibitem{DCS}
D. Baron, M. B. Wakin, M. F. Duarte, S. Sarvotham, and R. G. Baraniuk, 
\emph{Distributed Compressed Sensing}, Technical Report ECE-0612, 
Electrical and Computer Engineering Department, Rice University, December 2006. 
 \end{thebibliography}
\end{document}